\documentclass[conference]{IEEEtran}

\usepackage[T1]{fontenc}
\usepackage{graphicx}
\usepackage{cite}
\usepackage{subcaption}

\usepackage{overpic}
\usepackage{amssymb}
\usepackage{amsmath}
\usepackage{array}

\IEEEoverridecommandlockouts

\newtheorem{theorem}{Theorem}

\newtheorem{corollary}{Corollary}

\newcommand{\vect}[1]{\mathbf{#1}}
\def\tr{\mathrm{tr}}
\def\CN{\mathcal{N}_{\mathbb{C}}}

\begin{document}

\title{How Energy-Efficient Can a Wireless Communication System Become?}

\author{
	\IEEEauthorblockN{Emil Bj\"ornson and Erik G. Larsson \thanks{This paper was supported by ELLIIT and grants from the Swedish Foundation for Strategic Research (SSF) and the Swedish Research Council (VR).}}
	\IEEEauthorblockA{Department of Electrical Engineering (ISY), Link\"oping University, Sweden;
		\{emil.bjornson, erik.g.larsson\}@liu.se}
}

\maketitle

\begin{abstract}
The data traffic in wireless networks is steadily growing. The long-term trend follows Cooper's law, where the traffic is doubled every two-and-a-half year, and it will likely continue for decades to come. The data transmission is tightly connected with the energy consumption in the power amplifiers, transceiver hardware, and baseband processing. The relation is captured by the energy efficiency metric, measured in bit/Joule, which describes how much energy is consumed per correctly received information bit. While the data rate is fundamentally limited by the channel capacity, there is currently no clear understanding of how energy-efficient a communication system can become. Current research papers typically present values on the order of 10\,Mbit/Joule, while previous network generations seem to operate at energy efficiencies on the order of 10 kbit/Joule. Is this roughly as energy-efficient future systems (5G and beyond) can become, or are we still far from the physical limits? These questions are answered in this paper. We analyze a different cases representing potential future deployment and hardware characteristics.
\end{abstract}

\IEEEpeerreviewmaketitle

\section{Introduction}

A new wireless technology generation is introduced every decade and the standardization is guided by the International Telecommunication Union (ITU), which provides the minimum performance requirements. For example, 4G was designed to satisfy the IMT-Advanced requirements \cite{ITU-IMTA} on spectral efficiency, bandwidth, latency, and mobility. Similarly, the new 5G standard \cite{Parkvall2017a} is supposed to satisfy the minimum requirements of being an IMT-2020 radio interface \cite{ITU-IMT2020}. In addition to more stringent requirements in the  aforementioned four categories, a new metric has been included in \cite{ITU-IMT2020}: energy efficiency (EE). A basic definition of the EE is \cite{Kwon1986a,massivemimobook}
\begin{equation} \label{eq:EE-definition}
\mathrm{EE} \, [\textrm{bit/Joule}]= \frac{\textrm{Data rate [bit/s]}}{\textrm{Energy consumption [Joule/s]} }.
\end{equation}
This is a benefit-cost ratio and the energy consumption term includes transmit power and dissipation in the transceiver hardware and baseband processing \cite{Mammela2017a,massivemimobook}. A general concern is that higher data rates can only be achieved by consuming more energy;
if the EE is constant, then $100\times$ higher data rate in 5G is associated with a $100\times$ higher energy consumption. This is an environmental concern since wireless networks are generally not powered from renewable green sources. It is desirable to vastly increase the EE in 5G, but IMT-2020 provides no measurable targets for it, but claims that higher spectral efficiency will be sufficient. There are two main ways to improve the spectral efficiency: smaller cells \cite{Hoydis2011c,Mammela2017a} and massive multiple-input multiple-output (MIMO)\cite{Marzetta2010a,Rusek2013a}. The former gives substantially higher signal-to-noise ratios (SNRs) by reducing the propagation distances and the latter allows for spatial multiplexing of many users and/or higher SNRs. Since these gains are achieved by deploying more transceiver hardware per km$^2$, higher spectral efficiency will not necessarily improve the EE; the EE first grows with smaller cell sizes and more antennas, but there is an inflection point where it starts decaying instead \cite{Bjornson2015a}. The bandwidth is fixed in these prior works, but many other parameters are optimized for maximum EE. There are other non-trivial tradeoffs, such as the fact that transceiver hardware becomes more efficient with time \cite{Debaillie2015a,Mammela2017a}, so the energy consumption of a given network topology gradually reduces.

While the Shannon capacity \cite{Shannon1949b} manifests the maximal spectral efficiency over a channel and the speed of light limits the latency, the corresponding upper limit on the EE is unknown. A comprehensive study of the EE of 4G base stations is found in \cite{Auer2011a}. It shows that a macro site delivering 28\,Mbit/s has an energy consumption of 1.35\,kW, leading to an EE of 20\,kbit/Joule. Recent papers report EE numbers in the order of 10\,Mbit/Joule \cite{Venturino2015a,massivemimobook,Bjornson2016aabb} when considering future 5G deployment scenarios and using estimates of current transceivers' energy consumption. There is also numerous papers that consider normalized setups (e.g.,  1\,Hz of bandwidth) that give no insights into the EE that can be achieved in practice. Finally, the channel capacity per unit cost was studied for additive white Gaussian noise (AWGN) channels in \cite{Verdu1990a}, which is a rigorous but normalized form of EE analysis.

The goal of this paper is to analyze the physical EE limits in a few different cases and, particularly, give practically relevant numbers on the maximum achievable EE.

\section{An Ultimate Limit on the Energy Efficiency}

In this section, we derive an ultimate upper limit on the EE. We assume that the channels are deterministic and a consequence of this assumption is that perfect channel state information (CSI) is available everywhere (i.e., it can be estimated to any accuracy with a negligible overhead). Note that the capacity of a fading channel can be upper bounded by a deterministic channel having the channel realization from the fading distribution that maximizes the mutual information.\footnote{The conventional models for small-scale fading, including the Rayleigh fading model, have no upper bounds on the channel gain. However, any physical channel will have a finite-valued ``best'' realization since one cannot receive more power than was what transmitted; see Section~\ref{subsec:MIMO-case} for details.}

We will consider two cases: Single-antenna systems and multiple-antenna systems. 
In both cases, we assume that the communication takes place over a bandwidth of $B$\,Hz, the total transmit power is denoted by $P$\,W, and $N_0$\,W/Hz is the noise power spectral density. We treat $B$ and $P$ as design variables.

\subsection{Single-antenna Systems Without Interference}

\begin{figure}[t!]
	\centering \vspace{-2mm}
	\begin{overpic}[width=\columnwidth,tics=10]{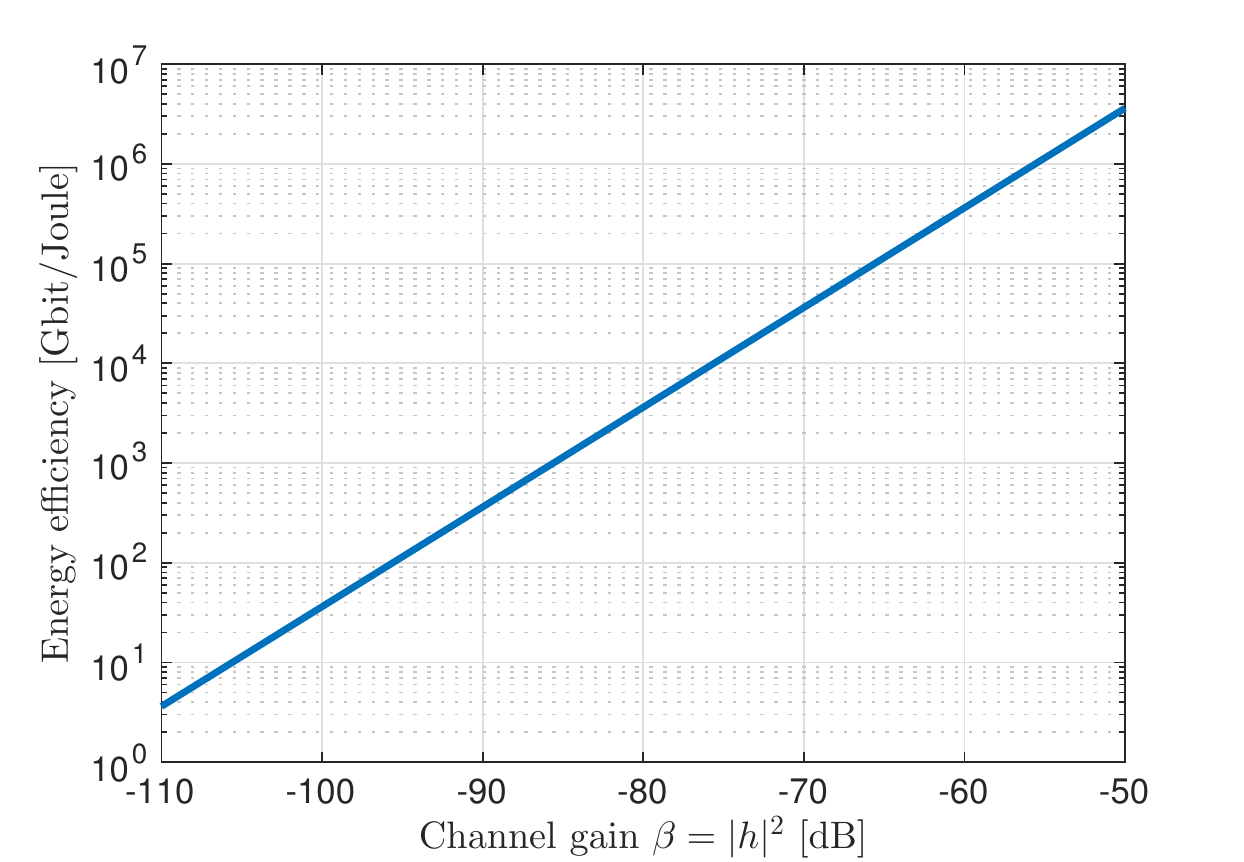}
 \put (75,44) {$8$\,m}
\put(77.6,47){\vector(0,1){5}}
 \put (48,27) {$80$\,m}
\put(51.75,30.6){\vector(0,1){5}}
 \put (22,27) {$800$\,m}
\put(25.75,26){\vector(0,-1){5}}
\end{overpic} 
	\caption{The maximum EE in a single-antenna system depends on the channel gain $\beta$. The propagation distances are computed for free-space propagation at 3\,GHz with lossless isotropic antennas, while the distances are often much shorter in practice.}
	\label{fig:simulationFigure_SISO} \vspace{-2mm}
\end{figure}
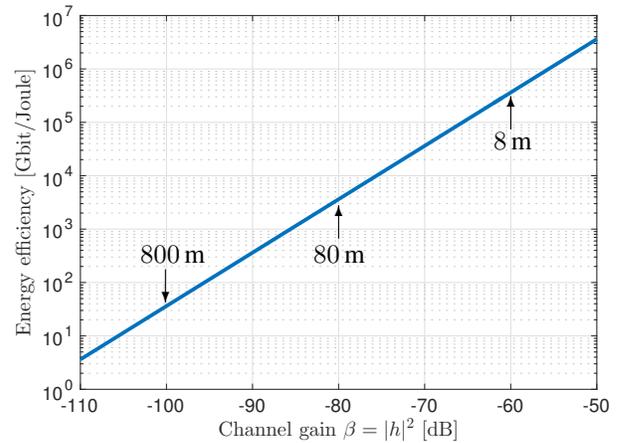

We begin by considering a single-antenna system. 
The channel is represented by a scalar coefficient $h \in \mathbb{C}$. The received signal $y \in \mathbb{C}$ is given by
\begin{equation}
{y} = h x +n
\end{equation}
where $x \in \mathbb{C}$ is the transmit signal with power $P$ and $n \sim \CN({0}, BN_0 )$ is AWGN. Since perfect CSI is available, the capacity of the channel is \cite{Shannon1949b}
\begin{equation} \label{eq:capacity-SISO}
C = B \log_2 \left( 1 + \frac{P \beta}{B N_0} \right) \quad \textrm{[bit/s]}
\end{equation}
where $\beta = |h|^2$ denotes the channel gain. The capacity is achieved by $x \sim \CN(0,P)$. When the transmit power is the only factor contributing to the energy consumption, an upper bound on the EE in \eqref{eq:EE-definition} is 
\begin{equation} \label{eq:EE-first-step-SISO}
 \frac{B \log_2 \left( 1 + \frac{P \beta}{B N_0} \right) }{P},
\end{equation}
which is a monotonically increasing function with respect to $B/P$. Hence, the EE is maximized as $P/B \to 0$, which can be achieved by taking the transmit power $P \to 0$, taking the bandwidth $B \to \infty$, or a combination thereof. 
The limit is easy to compute by considering a Taylor expansion of the logarithm around $\frac{P \beta}{B N_0}=0$: 
\begin{align} \notag
&\frac{B \log_2 \left( 1 + \frac{P \beta}{B N_0} \right) }{P} = \frac{B \log_2 (e) }{P} \!\! \left(\!\!  \frac{P \beta}{B N_0} - \underbrace{\sum_{n=2}^{\infty} (-1)^{n} \frac{\left( \! \frac{P \beta}{B N_0} \! \right)^{\!n}}{n}}_{\mathcal{O} \left( (\frac{P}{B})^2 \right) } \!\! \right) \\ & \to \frac{\log_2 (e) \, \beta}{N_0} \quad \textrm{as} \quad \frac{P}{B} \to 0, \label{eq:EE-SISO}
\end{align}
where $e$ denotes Euler's number. We recognize this as the reciprocal of the classical minimum energy-per-bit $N_0/\log_2(e) = N_0 \ln(2)$ for an AWGN channel \cite{Verdu1990a}, with the only difference that a deterministic channel gain $\beta$ has been included. To quantify the EE that can be achieved in this case, we use the typical noise power spectral density $N_0 = -174$\,dBm/Hz in room temperature and consider a practical range of channel gains $\beta$ from $-110$\,dB to $-50$\,dB.\footnote{The channel gain is seldom higher than $-50$\,dB. This value is achieved when communicating over 2.5\,m in the $3$\,GHz band  in free-space propagation, using lossless isotropic antennas. The value will decrease at higher carrier frequencies and when considering longer propagation distances.} The resulting EE is shown in Fig.~\ref{fig:simulationFigure_SISO} and ranges from $3$\,Gbit/Joule to $3\cdot 10^6$\,Gbit/Joule $= 3$ Pbit/Joule. These numbers are the EE limits in single-antenna systems with typical channel gains and are surely far from what is achieved by current systems.

\subsection{Single-antenna Systems With Interference}

We will now add interference to the system. The interference is caused by one or multiple systems that are also operating with maximum EE as goal. Hence, each transmitter uses the same transmit power $P$ and we let $\alpha>0$ denote the sum of the channel gains from all the interfering transmitters, leading to a total received interference  power of $P \alpha$. By treating interference as noise, the EE in \eqref{eq:EE-first-step-SISO} becomes
\begin{equation}
 \frac{B \log_2 \left( 1 + \frac{P \beta}{B N_0 + P \alpha} \right) }{P},
\end{equation}
which is still an increasing function of $B/P$. Hence, an upper bound on the EE is achieved by letting $P/B \to 0$, which leads to 
\begin{equation}
 \frac{\log_2(e) \beta}{N_0}.
\end{equation}
This expression is independent of $\alpha$ and, therefore, coincides with the limit in \eqref{eq:EE-SISO} for interference-free systems. This demonstrates that it was optimal to treat interference as noise in this case. Notice that we did not purposely neglect the interference, but the EE is maximized in the low SNR regime $P/B \to 0$ where the system is noise limited, not interference limited.

\subsection{Multiple-antenna Systems}
\label{subsec:MIMO-case}

Suppose the transmitter is equipped with $M$ antennas and the receiver is equipped with $N$ antennas, which is a MIMO system. The deterministic channel is now described by the channel matrix
$ \vect{H} \in \mathbb{C}^{N \times M}$. If we assume that there is no interference, the received signal $\vect{y} \in \mathbb{C}^N$ is
\begin{equation}
\vect{y} = \vect{H} \vect{x} + \vect{n}
\end{equation}
where $\vect{x} \in \mathbb{C}^M$ is the transmit signal and $\vect{n} \sim \CN(\vect{0}, BN_0 \vect{I}_N)$ is AWGN. The channel capacity of this MIMO system is \cite{Telatar1999a}
\begin{align}
C = \max_{\vect{K} \succeq \vect{0}:\, \tr( \vect{K} ) \leq P} B \log_2 \det \left( \vect{I}_N + \frac{1}{B N_0} \vect{H} \vect{K} \vect{H}^H \right)
\end{align}
and is achieved by $\vect{x} \sim \CN ( \vect{0}, \vect{K} )$ where the positive semi-definite correlation matrix $ \vect{K}$ is selected based on the waterfilling algorithm. An upper bound on the capacity is obtained when all the singular values of $\vect{H}$ are equal to the maximum singular value $\sigma_{\max}(\vect{H})$ of the matrix. We then obtain
\begin{align} 
C & \leq  \sum_{i=1}^{\min(M,N)} B \log_2 \left( 1+ \frac{P}{M B N_0} \sigma_{\max}^2(\vect{H}) \right) \notag \\
& = \min(M,N) B \log_2 \left( 1+ \frac{P}{M B N_0} \sigma_{\max}^2(\vect{H}) \right). \label{eq:MIMO-capacity-bound}
\end{align}
When the transmit power is the only factor contributing to the energy consumption, an upper bound on the EE in \eqref{eq:EE-definition} is 
\begin{align} \notag
&\frac{\min(M,N) B \log_2 \left( 1+ \frac{P}{M B N_0} \sigma_{\max}^2(\vect{H}) \right)}{P} \\
&\leq \frac{\min(M,N)  }{M} \frac{ \log_2 (e) \sigma_{\max}^2(\vect{H}) }{ N_0}  
\end{align}
where the upper limit is achieved by letting $P/B \to 0$ as in the single-antenna case.
The first term $\frac{\min(M,N)  }{M}$ is upper bounded by one and this bound is tight when the receiver has at least as many antennas as the transmitter.

\begin{figure}[t!]
	\centering
	\begin{overpic}[width=.7\columnwidth,tics=10]{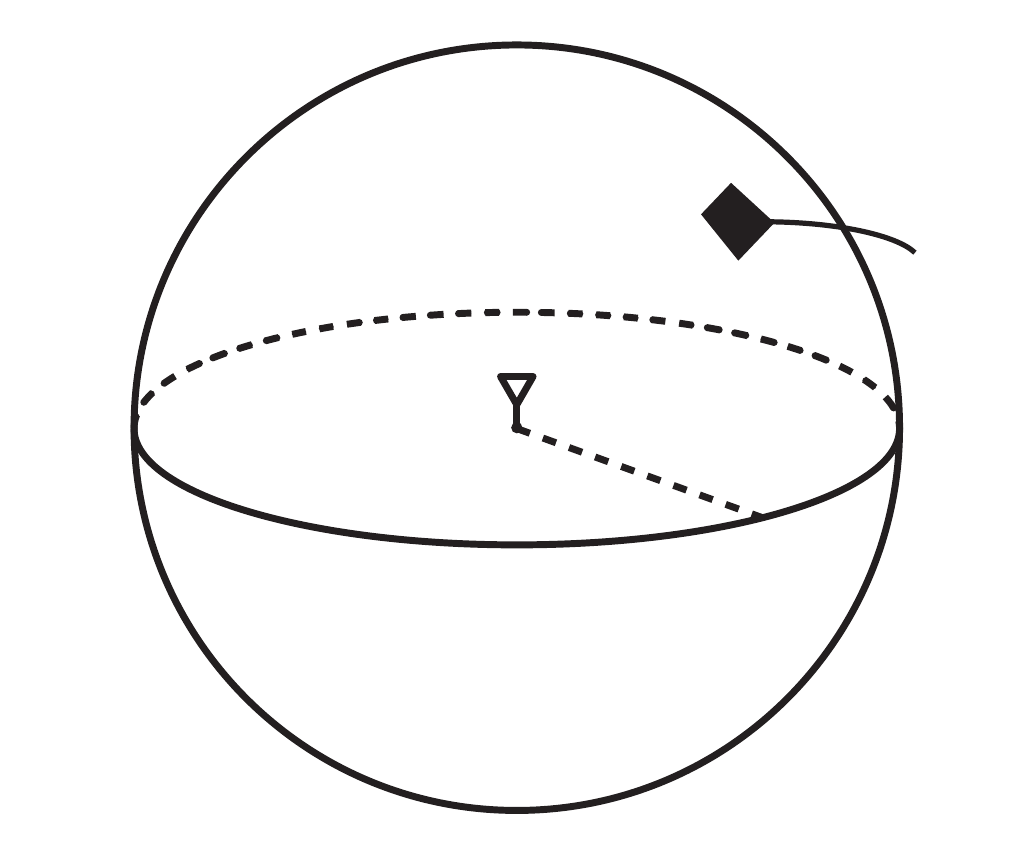}
 \put (62,40) {$r$}
  \put (91,58) {Receive}
    \put (91,52) {antenna}
     \put (91,46) {with area $A$}
\end{overpic} \vspace{-2mm}
	\caption{All the transmitted signal energy can be captured by enclosing the transmit antenna with a spheric surface of lossless receive antennas. We need $4\pi r^2 / A$ antennas to do so, where $A$ is the area of an antenna and $r$ is the radius.}
	\label{fig:sphere} \vspace{-2mm}
\end{figure}
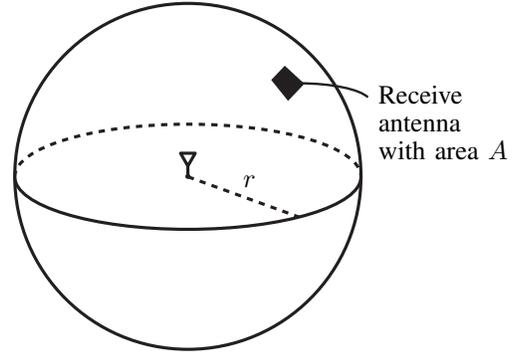

A more complicated question is how $\sigma_{\max}^2(\vect{H})$ depends on $M$ and $N$. Since we have assumed that all the non-zero singular values of $\vect{H}$ are equal, it follows that
\begin{equation}
\sigma_{\max}^2(\vect{H}) = \frac{\| \vect{H} \|_F^2}{\min(M,N)},
\end{equation}
where $\| \cdot \|_F$ is the Frobenius norm. Suppose all the elements of $\vect{H}$ have a constant magnitude $\sqrt{\beta}>0$, then $\sigma_{\max}^2(\vect{H}) = \beta MN / \min(M,N) = \beta \max(M,N)$, which goes to infinity as the number of transmit and/or receive antennas grow. This a common channel model in the Massive MIMO literature \cite{Marzetta2010a,Rusek2013a,Bjornson2015a,massivemimobook}, where it is utilized to demonstrate that the received signal power grows proportionally to the number of antennas. This scaling behavior makes sense for practical number of antennas, but not asymptotically; if the transmit power is $P$, the
law of conservation of energy manifests that the receiver can never receive more signal power than $P$, irrespectively of how many antennas are used. Hence, the physical upper limit on the singular values is one:
\begin{equation}
\sigma_{\max}^2(\vect{H}) \leq 1.
\end{equation}
The upper limit can be achieved by enclosing the transmitter by a sphere and then covering the surface of that sphere with  receive antennas. When the surface is fully covered, all the transmitted energy will be captured by the receive antennas, assuming that these are ideal (lossless). Suppose the sphere has radius $r$ and each lossless antennas has area $A$, as illustrated in Fig.~\ref{fig:sphere}, then we need $4\pi r^2 / A$ antennas to cover the surface. For example, if $r=10$\,m and isotropic antennas designed for the 3\,GHz band are used, then $A=0.1^2/(4\pi)$ and, consequently, we need 1.6 million antennas to cover the surface. This huge number explains why the asymptotic analysis in the Massive MIMO literature makes sense even in extreme practical cases with thousands of antennas. If we eventually want more than $4\pi r^2 / A$ antennas, we need to make the surface larger by increasing $r$. The consequence is that $\beta$ is reduced as $1/r^2$ and, therefore, we need to cover the 
larger surface of the new sphere with more antennas to capture the same energy.

In practice, we will most likely have $\sigma_{\max}^2(\vect{H}) \ll 1$, but we can set $\sigma_{\max}^2(\vect{H}) = 1$ to obtain the ultimate EE limit.
Hence, the EE of a multiple-antenna system is upper bounded as
\begin{equation} \label{eq:EE-MIMO}
\mathrm{EE} \leq  \frac{ \log_2 (e) }{ N_0}
\end{equation}
which is similar to the EE limit for single-antenna systems in \eqref{eq:EE-SISO}, but the key difference is that the channel gain $\beta$ has now been replaced with its upper bound: 0\,dB. If we insert the noise power spectral density into this expression, we achieve the ultimate EE limit: $10^{20.6}\,\textrm{bit/Joule} = 398$\,Ebit/Joule.

\section{Energy Efficiency Including Circuit Power}

The previous section demonstrated several ways to achieve high EE. The maximum is achieved when $P/B \to 0$. From an EE perspective, the analysis shows that it doesn't matter if $P\to 0$ or $B \to \infty$, but in terms of the data rate in \eqref{eq:capacity-SISO} it makes a huge difference:
\begin{equation}
C = B \log_2 \left( 1 + \frac{P \beta}{B N_0} \right) \to \begin{cases}
0, & P \to 0, \\
 \frac{ \log_2 (e) \, P \beta}{N_0}, & B \to \infty.
\end{cases}
\end{equation}
For example, we get $0$\,bit/s if $P \to 0$ or $1$\,Tbit/s if $B \to \infty$ (with $P = 20$\,dBm, $\beta = -75$\,dB, and $N_0 = -174$\,dBm/Hz). A communication system with zero capacity is practically worthless, even if it is energy-efficient from a purely mathematical perspective. One reason for this weird result is that we considered an energy consumption model where only the transmit power is included, but this will be generalized below.

\begin{figure}[t!]
	\centering \vspace{-2mm}
	\includegraphics[width=\columnwidth]{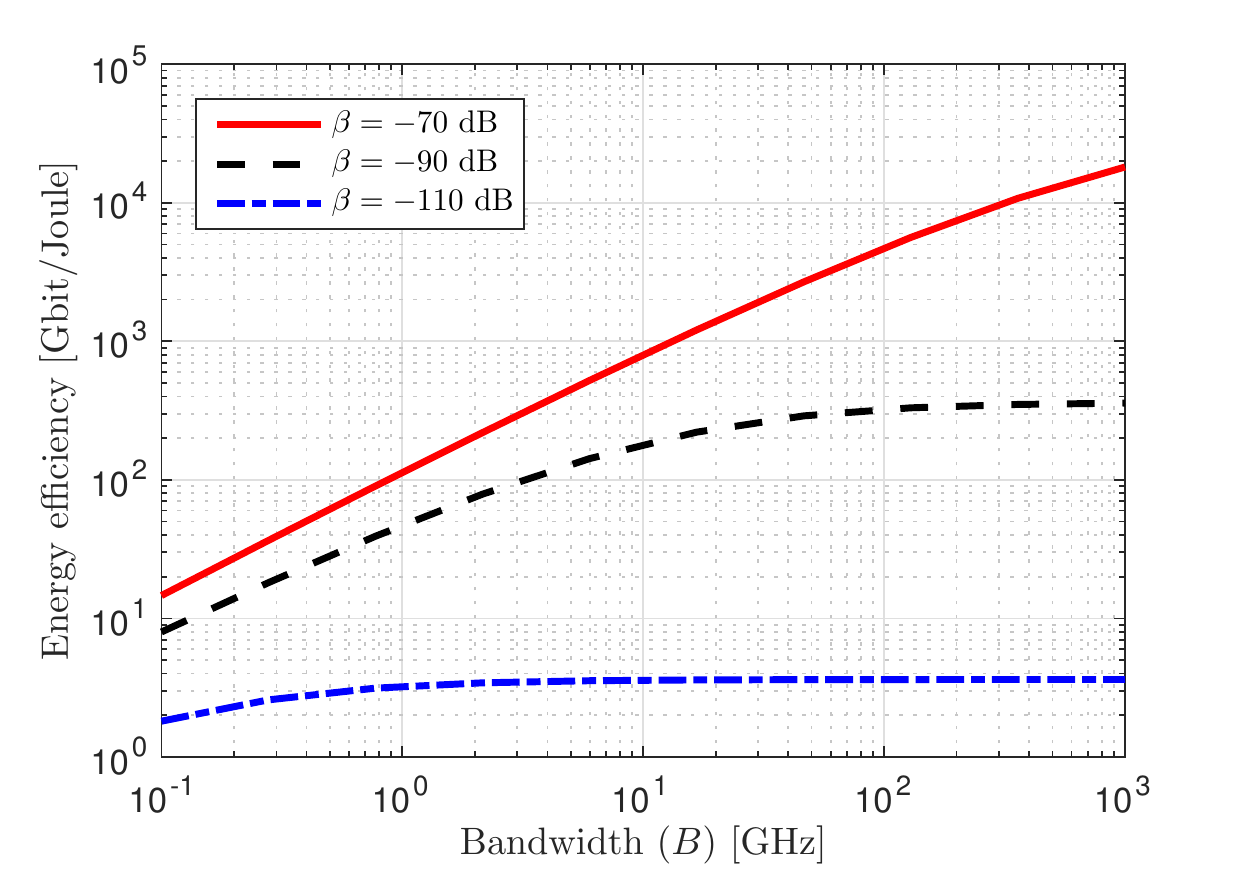}
	\caption{The EE increases with the bandwidth. The limit and the convergence depend strongly on the channel gain $\beta$.}
	\label{fig:simulationFigure_Bandwidth} \vspace{-2mm}
\end{figure}

Fig.~\ref{fig:simulationFigure_Bandwidth} shows how the EE approaches its limit as $B \to \infty$ when $P = 20$\,dBm and $N_0 = -174$\,dBm/Hz. Different values of $\beta$ are considered and these are determining how quickly we approach the EE limit. For the cell-edge case of $\beta = -110$\,dB, the limit is reached already at $B=1$\,GHz, while we need $100\times$ more bandwidth every time $\beta$ is increased by 20\,dB.

\subsection{Constant Circuit Power}

A more practical energy consumption model is $P + \mu$, where $\mu \geq 0$ is the circuit power---the power dissipated in the analog and digital circuitry of the transceivers. When communicating over long distances, it is common to have $P+\mu \approx P$, but in future smalls cells it is possible that $\mu> P$ \cite{Mammela2017a,Bjornson2016aabb}. 
In the single-antenna case without interference, the EE in \eqref{eq:EE-first-step-SISO} can now be generalized and upper bounded as
\begin{equation} \label{eq:EE-first-step-SISO-circuit-power}
\mathrm{EE} =  \frac{B \log_2 \left( 1 + \frac{P \beta}{B N_0} \right) }{P+\mu} \overset{(a)}{\leq}  \frac{\log_2 (e) \frac{P \beta}{ N_0}  }{P+\mu} \overset{(b)}{\leq}  \frac{ \log_2 (e) \beta}{ N_0},
\end{equation}
where $(a)$ follows from noting that the EE is an increasing function of $B$ and letting $B \to \infty$, while $(b)$ follows from letting $P \to \infty$. Another way to view it is that $P$ and $B$ are going jointly to infinity, but $B$ has a substantially higher convergence speed such that $P/B \to 0$.

Interestingly, the upper bound in \eqref{eq:EE-first-step-SISO-circuit-power} is the same as in \eqref{eq:EE-SISO}, thus the inclusion of the circuit power $\mu$ did not change the EE limit, but only made the conditions for achieving it stricter and more practical. Note that $\mu$ was not purposely removed in the bounding, but made negligible by taking $P \to \infty$.

\subsection{Varying Circuit Power}

The fact that we treated $ \mu$ as constant when changing $B$ and $P$ implies that no substantial changes to the hardware are needed when changing these variables. This simplification is hard to justify when taking the variables to infinity. The sampling rate is proportional to $B$ and the energy consumption of analog-to-digital and digital-to-analog converters is proportional to the sampling rate (i.e., behaves as $\nu B$ for some constant $\nu$), and the same applies to the baseband processing of these samples.
The energy consumption of data encoding/decoding is (at best) proportional to the data rate \cite{Schlegel2014a}. An alternative EE expression capturing these properties is
\begin{equation} \label{eq:EE-first-step-SISO-circuit-power2}
\mathrm{EE} =  \frac{B \log_2 \left( 1 + \frac{P \beta}{B N_0} \right) }{P + \nu B + \eta B  \log_2 \left( 1 + \frac{P \beta}{B N_0}  \right) } 
\end{equation}
where $\nu \geq 0$ and $\eta \geq 0$ are hardware-characterizing constants. 

\begin{theorem} \label{maximum-EE-lemma}
The EE in \eqref{eq:EE-first-step-SISO-circuit-power2} is maximized for any values of $P$ and $B$ such that
\begin{equation} \label{eq:optimal-ratio}
\frac{P}{B} = N_0 \frac{e^{x}-1}{\beta},
\end{equation}
where
\begin{equation}
x= W \left(\frac{\beta  \nu}{N_0 e} -\frac{1}{e}\right)+1
\end{equation}
and $W (\cdot)$ denotes the Lambert W function \cite{Bjornson2015a}.
\end{theorem}
\begin{IEEEproof}
With $z= P/B$,  \eqref{eq:EE-first-step-SISO-circuit-power2} can be expressed as
\begin{equation} \label{eq:EE-first-step-SISO-circuit-power2-z}
  \frac{ \log_2 \left( 1 + \frac{ \beta}{N_0} z \right) }{z + \nu + \eta \log_2 \left( 1 + \frac{ \beta}{ N_0} z \right) } 
\end{equation}
and can be directly maximized by using \cite[Lemma~3]{Bjornson2015a}.
\end{IEEEproof}

This theorem proves that the maximum EE is achieved when $P$ and $B$ have a non-zero finite ratio, in the practical case of $\nu >0$. The optimal ratio depends on the propagation condition (via $\beta/N_0$) and the transceiver hardware (via $\nu$). Interestingly, there is no dependence on $\eta$ which demonstrates that the energy consumption of the  encoding/decoding does not affect the optimal values of $P$ and $B$, but only the maximum value of the EE. Hence, it is the term $\nu B$ in \eqref{eq:EE-first-step-SISO-circuit-power2}  that fundamentally changes the behavior as compared to the previous subsections.

By inserting \eqref{eq:optimal-ratio} into \eqref{eq:EE-first-step-SISO-circuit-power2}, the maximum EE is obtained as
\begin{equation}
\mathrm{EE} =  \frac{   x  \log_2(e) }{N_0 \frac{e^{x}-1}{\beta} + \nu + \eta  x  \log_2(e) }.
\end{equation}
Since this EE is achieved by any values of $P$ and $B$ having the ratio in \eqref{eq:optimal-ratio}, we have the freedom to choose $B$ to achieve any desired data rate
\begin{equation}
C = B x \log_2(e).
\end{equation}
The corresponding EE-maximizing value of $P$ is obtained from Theorem~\ref{maximum-EE-lemma}. In other words, there is no tradeoff between EE and rate---except if $P$ and $B$ are limited by external factors.

\begin{figure*}[t!]
	\centering \vspace{-4mm}
\begin{subfigure}[b]{\columnwidth}
	\begin{overpic}[width=\columnwidth,tics=10]{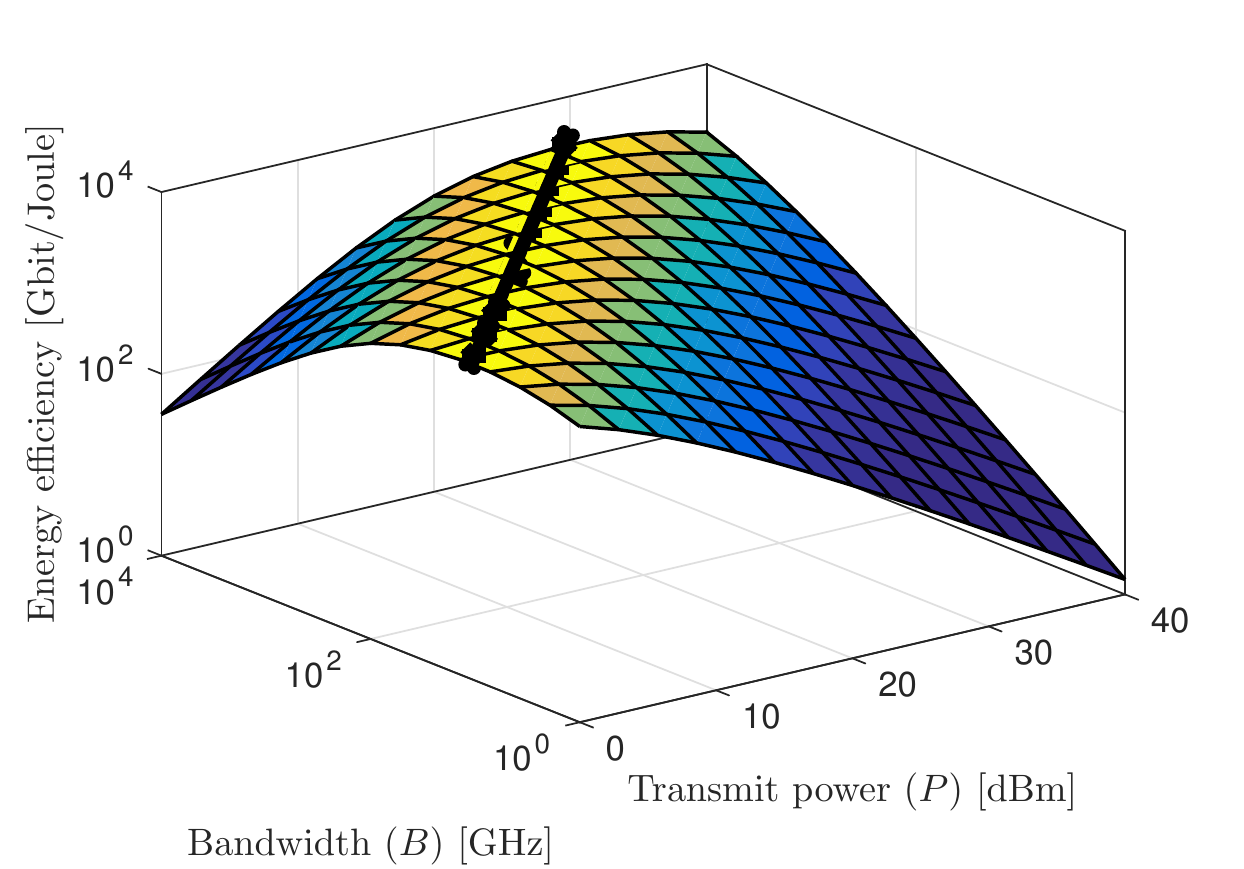}
  \put (17,32) {Maximum EE}
  \put(32,35){\vector(1,1){5}}
\end{overpic} \vspace{-8mm}
        \caption{}
    \end{subfigure}%
    \begin{subfigure}[b]{\columnwidth}
	\begin{overpic}[width=\columnwidth,tics=10]{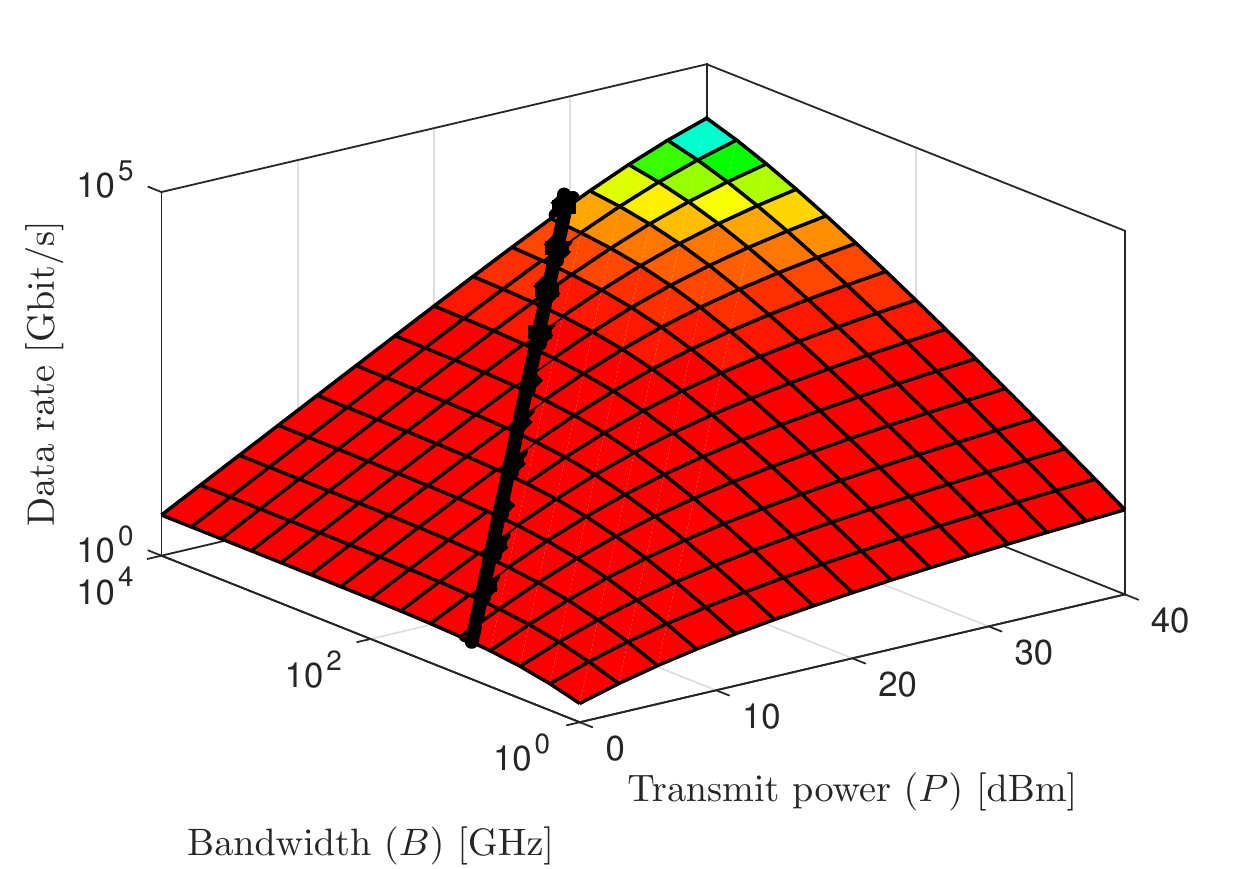}
  \put (15,51.5) {Maximum EE}
  \put(38.5,53){\vector(1,0){5}}
\end{overpic} \vspace{-8mm}
        \caption{}
    \end{subfigure} \vspace{-2mm}
    \caption{The EE in (a) and data rate in (b) vary with the transmit power and bandwidth. The maximum EE is achieved when the ratio in \eqref{eq:optimal-ratio} is fulfilled and we can then vary the power and bandwidth (along the thick line) to achieve any data rate needed.}         \label{fig:simulationFigureHardware} \vspace{-4mm}
\end{figure*}

These results are illustrated in Fig.~\ref{fig:simulationFigureHardware} for $\beta = -80$\,dB, $N_0 = -174$\,dBm/Hz, $\nu = 10^{-14}$\,J, and $\eta = 10^{-15}$\,J/bit. The latter two values are selected futuristically based on the fundamental bound on computing power \cite{Schlegel2014a,Mammela2017a}: the Landauer limit is approximately $10^{-18}$ logic operations per Joule. Hence, $\nu$ corresponds to 10000 logic operations per sample and $\eta$ to 1000 logic operations per bit.\footnote{One 16-bit multiplication requires around 3000 gates \cite{Mammela2017a}, thus the provided numbers correspond to very rudimentary processing and encoding/decoding.} Fig.~\ref{fig:simulationFigureHardware}(a) shows how the EE is maximized for certain combinations of $P$ and $B$, which are marked by a line. All these points provide the maximum EE of $3$\,Tbit/Joule, but they provide vastly different data rates, as shown in Fig.~\ref{fig:simulationFigureHardware}(b). In the considered parameter intervals, the EE-maximizing rate ranges from 0.3\,Gbit/s to 3\,Tbit/s. The EE-maximizing ratio $P/B$, provided by Theorem~\ref{maximum-EE-lemma}, gives an optimal SNR of $\frac{P\beta}{B N_0}=-6$\,dB and a spectral efficiency of 0.3\,bit/s/Hz. A binary modulation scheme with channel coding can achieve this bit/s/Hz in a practical implementation. For example, LDPC decoding can be implemented with $10^{-16}\,$J/bit \cite{Schlegel2014a}, which is below the considered value of $\eta$.

\subsection{Multiple-antenna Systems}

We can extend the analysis to cover MIMO systems. For brevity, we assume that both the transmitter and receiver are equipped with $M$ antennas. An achievable upper bound on the capacity is given in  \eqref{eq:MIMO-capacity-bound} and the corresponding EE is 
\begin{equation} \label{eq:EE-first-step-MIMO-circuit-power1}
\mathrm{EE} =  \frac{M B \log_2 \left( 1+ \frac{P}{M B N_0} \sigma_{\max}^2(\vect{H}) \right) }{P + \nu B M + \eta B M \log_2 \left( 1+ \frac{P}{M B N_0} \sigma_{\max}^2(\vect{H}) \right) },
\end{equation}
where the first term in the denominator is the total transmit power, the second term is the energy consumption of processing $M$ parallel signals at the transmitter and receiver, and the third term is the energy consumption of encoding/decoding.
\begin{corollary} 
The EE in \eqref{eq:EE-first-step-MIMO-circuit-power1} is maximized for any values of $P$ and $B$ such that
\begin{equation}
 \frac{P}{MB} = N_0 \frac{e^{\tilde{x}}-1}{\beta} \quad
\textrm{with} \,\,\, \tilde{x}= W \left(\frac{\sigma_{\max}^2(\vect{H})  \nu}{N_0 e} -\frac{1}{e}\right)+1.
\end{equation}
\end{corollary}
\begin{IEEEproof}
If we set $z=  \frac{P}{MB}$, the EE has the same structure as in \eqref{eq:EE-first-step-SISO-circuit-power2-z} and can be maximized as in Theorem~\ref{maximum-EE-lemma}.
\end{IEEEproof}

We can once again achieve maximum EE and any data rate, simultaneously. By adding more antennas we can increase that channel gain $\sigma_{\max}^2(\vect{H})$ towards 1. With the same $\nu$ and $\eta$ as in the previous subsection, the ultimate EE is 0.6\,Pbit/Joule.

\section{Conclusion}

The answer to the question ``\emph{How energy-efficient can a wireless communication system become?}'' depends strongly on which parameter values can be selected in practice and the energy consumption modeling. If it is modeled to capture the most essential hardware characteristics, the optimal EE is achieved for a particular ratio of the transmit power $P$ and bandwidth $B$, which typically corresponds to a low SNR. Any data rate can be achieved by jointly increasing $P$ and $B$ while keeping the optimal ratio. The physical upper limit on the EE is around 1\,Pbit/Joule. For practical number of antennas and channel gains, we can rather hope to reach EEs in the order of a few Tbit/Joule (as in Fig.~\ref{fig:simulationFigureHardware}) in future systems.

\bibliographystyle{IEEEtran}
\bibliography{IEEEabrv,refs}

\end{document}